\documentclass[aps,prd,superscriptaddress,nofootinbib,11pt]{revtex4}
\usepackage[english]{babel}
\usepackage[utf8]{inputenc}
\usepackage{graphicx}   
\usepackage{slashed}
\usepackage{epstopdf}
\usepackage{verbatim}   
\usepackage{color}      
\usepackage{subfigure}  
\usepackage{multirow}
\usepackage{hyperref}   
\usepackage{float}
\usepackage{epsfig,rotating}
\usepackage{amsmath,amssymb}
\usepackage{dsfont}
\usepackage{slashed}
\restylefloat{table}
\numberwithin{equation}{section}

\newcommand{\vx}{\vec{x}}
\newcommand{\vp}{\vec{p}}
\newcommand{\vq}{\vec{q}}
\newcommand{\vk}{\vec{k}}

\newcommand{\be}{\begin{equation}}
\newcommand{\ee}{\end{equation}}
\newcommand{\bea}{\begin{eqnarray}}
\newcommand{\eea}{\end{eqnarray}}

\newcommand{\ket}[1]{|#1\rangle}
\newcommand{\bra}[1]{\langle#1|}

\begin{document}
\title{Infrared dressing in real time: emergence of  anomalous dimensions.}

\author{Mudit Rai}
\email{MUR4@pitt.edu} \affiliation{Department of Physics and
Astronomy, University of Pittsburgh, Pittsburgh, PA 15260}
\author{Lisong Chen}
\email{lic114@pitt.edu} \affiliation{Department of Physics and
Astronomy, University of Pittsburgh, Pittsburgh, PA 15260}
\author{Daniel Boyanovsky}
\email{boyan@pitt.edu} \affiliation{Department of Physics and
Astronomy, University of Pittsburgh, Pittsburgh, PA 15260}

 \date{\today}

\begin{abstract}
We implement a   dynamical resummation method (DRM) as an extension of the dynamical renormalization group to study the  time evolution of infrared dressing in non-gauge theories. Super renormalizable and renormalizable models  feature infrared divergences   similar to those of a theory at a critical point, motivating a renormalization group improvement of the propagator that yields a power law decay of the survival probability $\propto t^{-\Delta}$. The (DRM) confirms this decay, yields the dressed state and determines that the anomalous dimension $\Delta$ is completely determined by the slope of the spectral density at threshold independent of the ultraviolet behavior, suggesting certain universality for infrared phenomena. The dressed state is an entangled state of the charged and massless quanta. The entanglement entropy is obtained by tracing over the unobserved massless quanta. Its time evolution is determined by the (DRM), it is infrared finite and describes the information flow from the initial single particle to the asymptotic multiparticle  dressed state. We show that effective field theories of massless axion-like particles coupled to fermion fields   do not feature infrared divergences, and provide a criterion for infrared divergences in effective field theories  valid for non-gauge theories up to one loop.

\end{abstract}


\maketitle

\section{Introduction}
The infrared behavior of interacting quantum field theories featuring massless fields has been of longstanding interest within the context of scattering amplitudes and  the S-matrix in gauge theories\cite{bn,lee,chung,kino,kibble,yennie,weinberg,kulish}. Infrared singularities associated with the emission and/or absorption of soft massless   quanta by charged fields has continued to be studied within the context of gauge theories in high energy physics\cite{lavelle,ein,furu,haya}, quantum coherence and infrared phenomena\cite{carney,zell,tomaras}, as well as    precision calculation of physical observables motivated by collider experiments\cite{schwartz1,finites,schwartz2}, but also of infrared aspects of  gravity\cite{strominger1,strominger2}.

Our main interest in the subject is motivated by the possibility that soft bremsstrahlung could yield an important mechanism for production of ultralight dark matter  particles  in an expanding cosmology. Motivated by this possibility, in this article we explore the consequences of infrared divergences associated with emission and absorption of soft massless quanta directly in real time in \emph{non-gauge theories} thereby bypassing the subtle aspects associated with gauge invariance, but addressing the main physics of the infrared behavior and the dynamics of dressing in real time. As for example in QED   the infrared singularities associated with charged single particle  states are a consequence of the single particle mass shell coinciding with the  multiparticle threshold.

The focus of  this article is restricted to the study of infrared divergences associated with the dressing of charged single particle states arising from absorption and emission of massless neutral quanta in non-gauge theories, not on the more overarching infrared aspects of the S-matrix in gauge theories explored in refs.\cite{bn}-\cite{schwartz2}.

Our main objective is to study the dynamics of dressing in real time, namely the time evolution of an initial state and the nature of the asymptotic many-particle state that emerges from the dressing of the charged single particle state by soft massless quanta of the neutral field in the asymptotic long time limit.

While we are ultimately interested in the cosmological applications, for which an S-matrix approach that relies on the infinite time limit is not the most useful framework to study time dependent phenomena, initiating this study of real time dressing dynamics in Minkowski space time may prove relevant for further understanding of infrared phenomena in gauge theories and gravity. Recently\cite{collins} a  re-evaluation of the Lehmann, Symanzik and Zimmermann reduction formula for asymptotic states beginning with a finite time analysis and proceeding to the infinite time limit has exhibited the subleties of this limit. The framework introduced in this article may provide complementary further insights into asymptotic theory in cases in which infrared divergences are present.

\vspace{1mm}

\textbf{Brief summary of main results:}

In this article we introduce a dynamical resummation method(DRM)\cite{boyazeno}, based on a generalization of the dynamical renormalization group\cite{gold,goldbook,boyvega}  to study the time evolution of initial states and the physics of soft dressing directly in real time.  In this article we  focus on various non-gauge theories that feature infrared singularities akin to those found in QED, hence it is possible that the results found in this study may prove a useful guide in gauge theories, and perhaps, in gravity\cite{strominger1,strominger2}. Our main results are the following:

\vspace{1mm}

\textbf{Models with infrared divergences:} We consider both a super renormalizable model and a renormalizable model of a charged field coupled to a massless field which while featuring very different ultraviolet behavior  exhibit the same infrared threshold singularities. We establish  a parallel between the infrared singularities of these models and those associated with a theory at a critical point\cite{amit,goldbook}. We do so by mapping  the behavior of the single particle propagator near threshold     to that of a critical Euclidean field theory at a fixed point. We then   implement a renormalization group (RG) resummation of the infrared behavior that leads to scaling with anomalous dimension. Performing a Fourier transform in time of   the (RG) resummed propagator reveals that the survival probability of a single particle state decays in time asymptotically as a power law with an anomalous dimension $\propto t^{-\Delta}$.
A dynamical resummation  method (DRM) is introduced that provides a resummation of self-energy corrections directly in real time\cite{boyazeno}. This method is manifestly unitary and directly related to the dynamical renormalization group\cite{gold,goldbook,boyvega} but extends it in significant ways: not only it reproduces the power law decay in time with anomalous dimension $\Delta$ which is shown to be determined by the derivative of the spectral density at threshold,  but also    yields a  physical description of the dynamics of dressing of the charged particle  by a  soft cloud of massless quanta.

\vspace{1mm}

\textbf{Universality:}
We find that the infrared divergence is a consequence of a linearly vanishing   spectral density at threshold with a finite slope $\Delta$. Implementing the (DRM) leads to the survival probability of the single particle state decaying at asymptotically long time as $\propto t^{-\Delta}$, reproducing the result from the (RG) improved propagator. The anomalous dimension $\Delta$ is completely determined by the slope of the spectral density at threshold. Therefore we interpret this behavior as a manifestation of universality, in the sense that   models that feature very different ultraviolet behavior but similar infrared threshold behavior    with spectral densities vanishing linearly at threshold, yield similar asymptotic dynamics. Obviously different models yield different values of the anomalous dimension $\Delta$, however, whatever the value of $\Delta$ all of these models feature an asymptotic survival probability $\propto t^{-\Delta}$ with scaling behavior. This is similar to universality in critical phenomena where scaling behavior near a critical point is described in terms of critical exponents which are insensitive to the ultraviolet behavior of the theory.

\textbf{Massless axion-like particles:} Motivated by their possible relevance in cosmology, we studied the case of effective field theories of a massless axion-like pseudoscalar particle coupled to fermionic degrees of freedom. We considered both   pseudoscalar   and  pseudovector couplings. In both cases we find that the emission and absorption of the massless quanta results in spectral densities that vanish faster than linear at threshold, thus preventing infrared divergences. These theories do not feature decay with anomalous dimensions ($\Delta =0$). We provide a criterion for the determination of infrared divergences in general non-gauge effective field theories valid up to one loop level.

\textbf{The entangled dressing cloud and its entropy:} The (DRM) describes unitary time evolution and yields the asymptotic multiparticle state that results from the evolution of the initial single particle state. We show explicitly how unitarity is manifest in the asymptotic long time limit when the initial state has completely decayed (with a power law).  This asymptotic pure state is an entangled state    between the charged particle and the soft cloud with amplitudes that exhibit the infrared enhancement and the anomalous dimension.  If a detector only measures the charge of the asymptotic state, but is insensitive to the massless quanta, tracing the asymptotic state over the unobserved degrees of freedom yields a mixed state. The probabilities display the infrared enhancement, which is, however, compensated by  contributions vanishing with the anomalous dimension. The entanglement entropy is obtained directly in real time, its time evolution is completely determined by the (DRM) equations, it  describes the information flow from an initial single particle state to the asymptotic entangled multiparticle state and is infrared finite as a consequence of the anomalous dimension.

\section{Super renormalizable, and renormalizable  models:}\label{sec:IRmodels}
We study the dynamics of infrared dressing in two  models that feature different ultraviolet behavior but share similar infrared behavior near the multiparticle threshold, and effective field theory models of a charged fermion field coupled to a massless axion-like particle.
\subsection{Super renormalizable case:}\label{subsec:super}
Let us consider the case of a massive complex, charged scalar field $\phi$ coupled to a massless real scalar field $\chi$.
\be
\mathcal{L} = \partial^\mu \phi^\dagger \partial_\mu \phi - M^2 \phi^\dagger \phi +
\frac{1}{2}\,\partial^\mu \chi  \partial_\mu \chi- \lambda \,\phi^\dagger \phi \, \chi \label{lagsuper}\ee

Including the one-loop self energy, the Dyson-resummed $\phi$ propagator is
\be G_{\phi}(P) = \frac{1}{P^2 - M^2 - \Sigma(P^2)} \label{fiprop} \ee where
\be \Sigma(P^2) = -\frac{\widetilde{\lambda}^2}{(4\pi)^2}\,L + \frac{\widetilde{\lambda}^2}{(4\pi)^2}\,I(P^2/M^2) \label{sigsup1} \ee
where in dimensional regularization in dimension $D= 4-\varepsilon$
\be \widetilde{\lambda} = \lambda\,\mu^{-\varepsilon/2}~~;~~ L = \frac{2}{\varepsilon}-\gamma_E + \ln(4\pi) - \ln\Big[\frac{M^2}{\mu^2} \Big] \label{msbar} \ee and
\be I(\alpha) = \int^1_0 \ln\Big[x-\alpha\,x\,(1-x)-i\tilde{\epsilon} \Big]\, dx ~~;~~ \tilde{\epsilon} \rightarrow 0^+ \,. \label{Iofalfa} \ee Subtracting the self-energy at $P^2 = M^2_p$, the renormalized mass, at which the inverse propagator vanishes, namely
\be \Sigma(P^2) = \Sigma(M^2_p) + \overline{\Sigma}(P^2)\,, \label{subt}\ee where
\be M^2_p = M^2 + \Sigma(P^2=M^2_p) \label{pole} \ee it follows that
\be G(P^2) = \frac{1}{P^2-M^2_p - \overline{\Sigma}(P^2)} \,, \label{renprop}\ee with
\be \overline{\Sigma}(P^2) = \frac{\lambda^2_R}{(4\pi)^2}\, \frac{P^2-M^2_p}{P^2}\, \ln\Big[\frac{M^2_p - P^2 -i\tilde{\epsilon}}{M^2_p} \Big]\,. \label{oversigsup} \ee
To leading order we have replaced  bare by renormalized quantities in $\overline{\Sigma}$. Although the inverse propagator vanishes at $P^2 = M^2_p$, $d \overline{\Sigma}/dP^2$ features an infrared singularity at $P^2 = M^2_p$ which is the beginning of the multiparticle cut and the threshold for emission of soft quanta, since
\be \mathrm{Im}\overline{\Sigma}(P^2) = -\pi \,\frac{\lambda^2_p}{(4\pi)^2}\, \frac{P^2-M^2_p}{P^2}\,\,\Theta(P^2-M^2_p) \,.\label{imsigsup} \ee

Near $P^2 = M^2_p$ the propagator becomes
\be G(P^2) = \frac{1}{(P^2-M^2_p)\Big[1-g^2\,\ln\Big(\frac{M^2_p - P^2 }{M^2_p} \Big) \Big]} ~~;~~ g = \frac{\lambda_R}{4\pi\,M_p} \,, \label{Gnearpole} \ee where in the argument of the logarithm $P^2 \rightarrow P^2 + i\tilde{\epsilon}$. This behavior for $P^2 \simeq M^2_p$ is reminiscent of critical phenomena\cite{amit} which suggests the implementation of a renormalization group resummation, the details of which are presented in appendix (\ref{app:RG}). The result is  the renormalization group improved propagator
\be G^{RG}(P^2) = \frac{1}{(P^2-M^2_p)\Big[\frac{M^2_p - P^2 }{M^2_p} \Big]^{-g^2}}\,.
\label{RGprop} \ee The forward time evolution is obtained from the inverse Fourier transform in frequency:
\be \widetilde{G}(t) = \int \frac{dp_0}{2\pi} e^{-ip_0 t}\,G^{RG}(P^2)\,,  \label{tivol}\ee and the long time limit is determined by the behavior of $G^{RG}(P^2)$ for $P^2 \simeq M^2_p$. Writing $P^2 = (p_0-E_{p})(p_0+E_{p})$ with $E^2_{p} = \vp^2 + M^2_p$, and changing variables to $(p_0-E_{p})= x/t$ the integral becomes (with $x \rightarrow x + i\tilde{\epsilon}$)
\be \widetilde{G}(t) = e^{-iE_p t}\, \int \frac{dx}{2\pi} \frac{e^{-ix}}{x}\, \frac{i}{2E_p + \frac{x}{t}}\,\frac{1}{\Big[-\frac{x}{M_p\,t}\,\Big(\frac{2E_p+x/t}{M_p} \Big) \Big]^{-g^2}}\,, \label{Goft1}\ee which in the long time limit becomes
\be \widetilde{G}(t) \propto \frac{e^{-iE_p t}}{2E_p}\, \Big[ \frac{M_p}{2E_p}\big]^{-2g^2} \Big[ E_p \, t\Big]^{-g^2} \,.\label{ltG}\ee Therefore a renormalization group improvement of the branch cut singularity beginning at $P^2 = M^2_p$ yields  long time power law decay with anomalous dimension $g^2 = (\lambda_R/4\pi M_p)^2$. This asymptotic scaling behavior is a consequence of the infrared singularity at threshold of the propagator. The propagator (\ref{Goft1}) describes the asymptotic time evolution of the amplitude, therefore, the survival probability of  the initial state is
\be |\widetilde{G}(t)|^2  \propto \Big[ E_p\, t\Big]^{-2\,g^2}\,. \label{proba1} \ee For a typical decaying state this survival probability would be of the form $e^{-\Gamma t}$ with $\Gamma$ being the total decay width.

\subsection{Renormalizable case:} As an example of a renormalizable case we consider a Dirac fermion Yukawa coupled to a massless real scalar field $\Phi$, namely
\be \mathcal{L} = \frac{1}{2} \,\partial_{\mu} \Phi \partial^{\mu}\Phi  + \overline{\Psi} \Big ( i {\not\!{\partial}}-M \Big)\Psi - Y\,\overline{\Psi}\Phi \Psi\,. \label{yukaL}\ee The fermion propagator is given by
\be S(P) = \frac{i}{{\not\!{P}}-M-\Sigma(\slashed{P})} \,. \label{ferprop}\ee The one loop self energy is given by
\be \Sigma(\slashed{P}) = -\frac{\widetilde{Y}^2}{(4\pi)^2}\, \Bigg\{\Big(\frac{{\not\!{P}}}{2}+M\Big) L -\int^1_0 \big[{\not\!{P}}(1-x)+M \big]\ln\big[x-\alpha x(1-x)-i\tilde{\epsilon} \big]\,dx \Bigg\} \,,\label{ferself}\ee where in dimensional regularization with $D=4-\varepsilon$
\be \widetilde{Y}^2 = Y^2 \,\mu^{-\varepsilon}~~;~~ L = \Big\{\frac{2}{\varepsilon} - \gamma_E + \ln(4\pi) -\ln\big[\frac{M^2}{\mu^2}\big] \Big\} ~~;~~ \alpha = \frac{P^2}{M^2} \,. \label{Leq}\ee First, we renormalize the mass by requesting that the inverse propagator vanishes at ${\not\!{P}} = M_p$ from which it follows that
\be M_p = M + \Sigma({\not\!{P}} = M_p)\,, \label{polemass}\ee
secondly, we introduce the off-shell wave function renormalization constant $Z$ and renormalized coupling $y_R$ as
\be Z^{-1} = 1- \frac{\widetilde{Y}^2\,L}{2\,(4\pi)^2}  ~~;~~ y^2_R = \frac{Z\, \widetilde{Y}^2}{(4\pi)^2} \,\label{ZfergR}\ee yielding
\be S(\slashed{P}) = \frac{i \,Z}{{\not\!{P}}-M_p-\widetilde{\Sigma}(\slashed{P})} \,,\label{propa2} \ee where to leading order in the Yukawa coupling,
\be \widetilde{\Sigma}(\slashed{P}) = y^2_R \Big[{\not\!{P}}\, \Big(\frac{\alpha^2-1}{2\alpha^2}\Big)+M_p \,\Big(\frac{\alpha-1}{\alpha} \Big) \Big]\,\ln\Big[1-\alpha \Big]~~;~~ \alpha \equiv \frac{P^2}{M^2_p} + i\tilde{\epsilon} \,. \label{sigsubs}  \ee Near the mass shell ${\not\!{P}} \simeq M_p$ we find the behavior
\be S(\slashed{P}) =  \frac{i \,Z\, \Big({\not\!{P}}+M_p\Big)}{\Big(P^2-M^2_p\Big)\,\Big[1-4\,y^2_R \,\ln\Big[\frac{M^2_p-P^2}{M^2_p} \Big] \Big]}\,. \label{nearpole} \ee Up to the overall (ultraviolet divergent) constant $Z$ this propagator features the same type of infrared singularity as in the super renormalizable case and we invoke a similar renormalization group resummation (see appendix (\ref{app:RG})) leading to the renormalization group improved propagator
\be S^{RG}(P) = \frac{i \,Z\,\Big({\not\!{P}}+M_p\Big)}{\Big(P^2-M^2_p\Big)\, \Big[\frac{M^2_p-P^2}{M^2_p} \Big]^{-4y^2_R}  } ~~;~~ P^2 \rightarrow P^2 +i\tilde{\epsilon}  \, \,.\label{RGSf} \ee We note that the behavior near $P^2\simeq M^2_P$ is very similar to the super renormalizable case, given by eqn. (\ref{RGprop}).

As in the super renormalizable case, the forward time evolution is obtained by the inverse Fourier transform,   the long time limit is determined by the threshold region $P^2 \simeq M^2_P$. Projecting onto a positive energy spinor (for forward time evolution) and proceeding as in the previous case, with $p_0 - E_p = x/t$,
we find in the long time limit
\be \widetilde{S}(t) \propto  Z\,  e^{-iE_p t}\, \Big[\frac{M_R}{2E_p}\big]^{1-8y^2_R} \Big[ E_p\, t\Big]^{-4y^2_R} \,, \label{Soft1posE}\ee Again the scaling behavior at long time is a manifestation of the infrared singularity at threshold.

We note that in the form of the propagators (\ref{Gnearpole},\ref{nearpole}) the discontinuity of the propagator across the two particle cut vanishes linearly in $p_0-E_p$, this feature will prove to be important in the emergence of power law decay in time as explicitly shown by the  dynamical resummation method of next section.

\subsection{Massless axion-like particles:}\label{subsec:axions}

We consider massless axion-like particles as pseudoscalar real massless fields, and two different couplings to a  fermion field: \textbf{a:)}  pseudoscalar Yukawa coupling $ ig \phi\, \overline{\Psi} \gamma^5 \Psi$, \textbf{b:)} pseudovector   coupling $g \partial_{\mu} \phi \,\overline{\Psi}\gamma^{\mu}\gamma^5 \Psi$.

\vspace{1mm}

\textbf{a:)} This is also a renormalizable case. The propagator is given by eqn. (\ref{ferself}), and in this case, it is straightforward to conclude that the self-energy $\Sigma(\slashed{P})$ is obtained from that of the scalar case (\ref{yukaL}) by simply replacing $M \rightarrow -M$. Following the same renormalization procedure as in the scalar case, after mass and (off-shell) wave function renormalization the propagator in this, pseudoscalar case (ps), reads
\be S_{ps}(\slashed{P}) = \frac{i \,Z_{ps}}{{\not\!{P}}-M_p-\widetilde{\Sigma}_{ps}(\slashed{P})} \,,\label{propa2pseudo} \ee with

\be \widetilde{\Sigma}_{ps}(\slashed{P}) = g^2_R \Big[{\not\!{P}}\, \Big(\frac{\alpha^2-1}{2\alpha^2}\Big)- M_p \,\Big(\frac{\alpha-1}{\alpha} \Big) \Big]\,\ln\Big[1-\alpha \Big]~~;~~ \alpha \equiv \frac{P^2}{M^2_p} + i\tilde{\epsilon} ~~;~~ g^2_R = Z_{ps}\,\frac{g^2}{(4\pi)^2} \,. \label{sigpseudo}  \ee In this case, we now find that near the mass shell at $\slashed{P} \simeq M_p$ the propagator is
\be S_{ps}(\slashed{P}) =  \frac{i \,Z_{ps}\, \Big({\not\!{P}}+M_p\Big)}{\Big(P^2-M^2_p\Big)\,\Big[1- g^2_R\,\frac{(P^2-M^2_p)}{M^2_p} \,\ln\Big[\frac{M^2_p-P^2-i\widetilde{\epsilon}}{M^2_p} \Big] \Big]}\,. \label{nearpolepseudo} \ee The logarithm associated with the two-particle cut yields a contribution to the self-energy of the form
\be \Sigma_{cut}(P^2) \propto \frac{(P^2-M^2_p)^2}{M^2_p} \,\ln\Big[\frac{M^2_p-P^2-i\widetilde{\epsilon}}{M^2_p} \Big]\,,\label{sigcutps}\ee yielding
\be \mathrm{Im}\Sigma_{cut}(P^2) \propto \frac{(P^2-M^2_p)^2}{M^2_p} \Theta(P^2-M^2_p)\,, \label{imcut1}\ee

therefore the pseudoscalar axion coupling does not lead to infrared divergences.

\vspace{1mm}

\textbf{b:)} This is a non-renormalizable coupling, with $g$ featuring mass dimension $(-1)$ in 4-space-time dimensions. The one loop self energy is given by ($D=4-\varepsilon$)
\begin{align}
\Sigma(\slashed{P})&=-ig^2\int\frac{d^Dk}{(2\pi)^D}\frac{\slashed{k}\gamma^5(\slashed{P}+\slashed{k}+M)\slashed{k}\gamma^5}{k^2((k+P)^2-M^2)}\\
&=-ig^2\int\frac{d^Dk}{(2\pi)^D}\frac{k^2\slashed{k}-M k^2+2\slashed{k}k\cdot P-k^2\slashed{P}}{k^2((k+P)^2-M^2)}.
\end{align}
 which can be written as
\begin{equation}
 \Sigma(\slashed{P})= \slashed{P}\,\Sigma_V(P^2)+M\,\Sigma_S(P^2)\,.\label{decom}
\end{equation} with
\begin{align}
\Sigma_V(P^2)&= {g}^2\frac{(P^2+M^2)A_0 -(P^2-M^2)^2B_0(P^2)}{32\pi^2 P^2}\;\\
\Sigma_S(P^2)&= {g}^2\frac{A_0}{16\pi^2}\,
\end{align}
 and
\be
A_0 = M^2\big[ 1 + L \big]~~;~~  B_0(P^2) =2+L +\frac{M^2-P^2}{p^2}\ln\Big[\frac{M^2 -P^2-i\epsilon}{M^2}\Big]\,, \label{ABcero}\ee where $L$ is given by eqn. (\ref{Leq}). Although the divergence proportional to $\slashed{P}$  cannot be renormalized, it is clear  however, that near the mass-shell $P^2 \simeq M^2_p $ the logarithm describing the two particle cut yields a term of the form
\be \Sigma_{cut}(\slashed{P} \simeq M_p) \propto (P^2-M^2_p)^3\,\ln\Big[\frac{M^2_p-P^2-i\epsilon}{M^2_p}\Big]\,, \label{dercoupir}\ee yielding
 \be \mathrm{Im}\Sigma_{cut}(\slashed{P} \simeq M_p) \propto (P^2-M^2_p)^3 \Theta(P^2-M^2_p) \label{imcut2}\,,\ee
 therefore, also in this case there is no infrared divergence at the position of the mass-shell. Perhaps in this case this is an expected   consequence of the derivative coupling, which brings two extra powers of momenta in the loop that relieves  the infrared divergence.

 We conclude that in both cases,   either in the pseudoscalar or pseudovector axion coupling, there is no infrared divergence associated with the beginning of the multiparticle cut, at least up to   one-loop order studied here.

\section{\label{sec:WW} Dynamical resummation method (DRM):}
We now introduce a method that implements a dynamical resummation directly in time\cite{boyazeno} that is intimately related to the dynamical renormalization group\cite{gold,goldbook,boyvega}. We first describe the resummation method in generality, relate it to the dynamical renormalization group, and apply the results to the cases studied in the previous section.

Consider a system whose Hamiltonian $H=H_0+H_I$ with  $H_I$ a perturbation. The time evolution of states in the interaction picture
of $H_0$ is given by
\be i \frac{d}{dt}|\Psi_I(t)\rangle   = H_I(t)\,|\Psi_I(t)\rangle \,,  \label{intpic}\ee
where the interaction Hamiltonian in the interaction picture is
\be H_I(t) = e^{iH_0\,t} H_I e^{-iH_0\,t} \label{HIoft}\ee

This has the formal solution
\be |\Psi_I(t)\rangle  = U(t,t_0) |\Psi_I(t_0)\rangle  \label{sol}\ee
where   the time evolution operator in the interaction picture $U(t,t_0)$ obeys \be i \frac{d}{dt}U(t,t_0)  = H_I(t)U(t,t_0)\,. \label{Ut}\ee

Now we can expand \be |\Psi_I(t)\rangle  = \sum_n C_n(t) |n\rangle \,, \label{expan}\ee where $|n\rangle$ are eigenstates of the unperturbed Hamiltonian, $H_0 \ket{n} = E_n\,\ket{n}$, and form a complete set of orthonormal states. In the quantum field theory case these are  many-particle Fock states. From eq.(\ref{intpic}) one finds the {\em exact} equation of motion for the coefficients $C_n(t)$, namely

\be \dot{C}_n(t) = -i \sum_m C_m(t) \langle n|H_I(t)|m\rangle \,. \label{eofm}\ee

\begin{figure}[ht!]
\begin{center}
\includegraphics[height=3in,width=3in,keepaspectratio=true]{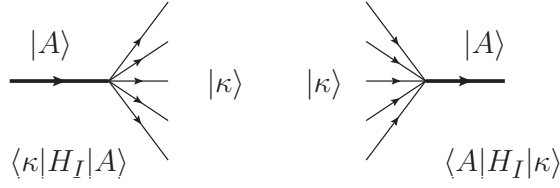}
\caption{Transitions $|A\rangle \leftrightarrow |\kappa\rangle$ in first order in $H_I$.}
\label{fig1:coupling}
\end{center}
\end{figure}

Although this equation is exact, it generates an infinite hierarchy of simultaneous equations when the Hilbert space of states spanned by $\{|n\rangle\}$ is infinite dimensional. However, this hierarchy can be truncated by considering the transition between states connected by the interaction Hamiltonian at a given order in $H_I$. Thus
consider the situation depicted in figure~\ref{fig1:coupling} where one state, $|A\rangle$, couples to a set of states $\left\{|\kappa\rangle\right\}$, which couple back  to $|A\rangle$ via $H_I$.

Under these circumstances, we have \bea \dot{C}_A(t) & = & -i \sum_{\kappa} \langle A|H_I(t)|\kappa\rangle \,C_\kappa(t)\label{CA}\\
\dot{C}_{\kappa}(t) & = & -i \, C_A(t) \langle \kappa|H_I(t) |A\rangle \label{Ckapas}\eea where the sum over $\kappa$ is over all the intermediate states coupled to $|A\rangle$ via $H_I$.

Consider the initial value problem in which at time $t_0=0$ the state of the system $|\Psi(t=0)\rangle = |A\rangle$ i.e. \be C_A(0)= 1,\   C_{\kappa}(0) =0 .\label{initial}\ee  We can solve eq.(\ref{Ckapas}) and then use the solution in eq.(\ref{CA}) to find
\bea  C_{\kappa}(t) & = &  -i \,\int_0^t \langle \kappa |H_I(t')|A\rangle \,C_A(t')\,dt' \label{Ckapasol}\\ \dot{C}_A(t) & = & - \int^t_0 \Sigma(t,t') \, C_A(t')\,dt' \label{intdiff} \eea where \be \Sigma(t,t') = \sum_\kappa \langle A|H_I(t)|\kappa\rangle \langle \kappa|H_I(t')|A\rangle = \sum_\kappa e^{i(E_A- E_\kappa)(t-t')}\,|\bra{A}H_I(0)\ket{\kappa}|^2\, \, \label{sigma} \ee where we used (\ref{HIoft}). It is convenient to write $\Sigma(t,t') $ in a spectral representation, namely
\be \Sigma(t,t') = \int^{\infty}_{-\infty}\,\rho(p_0)\,e^{-i(p_0-E_A)(t-t')}\,  dp_0 \label{specrep}\,,\ee where we have introduced the spectral density
\be \rho(p_0) = \sum_\kappa  \,|\bra{A}H_I(0)\ket{\kappa}|^2\,\delta(p_0-E_{\kappa})\,.  \label{rhopo}\ee

The  integro-differential equation  with {\em memory} (\ref{intdiff}) yields a non-perturbative solution for the time evolution of the amplitudes and probabilities. Inserting the solution for $C_A(t)$ into eq.(\ref{Ckapasol}) one obtains the time evolution of amplitudes $C_{\kappa}(t)$ from which we can compute  the time dependent probability to populate the state $|\kappa\rangle$, namely $|C_\kappa(t)|^2$.

The hermiticity of the interaction Hamiltonian $H_I$, and the equations (\ref{CA},\ref{Ckapas}) yield
\be \frac{d}{dt} \Big[|C_A(t)|^2 + \sum_{\kappa}|C_{\kappa}(t)|^2  \Big] = 0 \label{derit}\ee which  together with the initial conditions in eqs.(\ref{initial}) yields the unitarity relation
\be |C_A(t)|^2 + \sum_{\kappa} |C_{\kappa}(t)|^2 =1\,, \label{unitarity1}\ee which  is the statement that the time evolution operator $U(t,0)$ is unitary, namely
\bea \langle \Psi_I(t)|\Psi_I(t)\rangle  & = & |C_A(t)|^2 + \sum_{\kappa} |C_{\kappa}(t)|^2   \nonumber \\
 & =  & \langle \Psi(0)U^{\dagger}(t,0)   U(t,0) \Psi(0) \rangle = \langle \Psi(0)| \Psi(0) \rangle =|C_A(0)|^2 =1 \,.  \label{unistat}\eea

In general it is quite difficult to solve eq.(\ref{intdiff}) exactly, so that an approximation scheme must be invoked.

The time evolution of $C_A(t)$ determined by eq.(\ref{intdiff}) is \emph{slow} in the sense that
the time scale is determined by a weak coupling kernel $\Sigma$ which is second order in the coupling. This allows us to use an approximation in terms of a
consistent expansion in time derivatives of $C_A$. Define \be W_0(t,t') = \int^{t'}_0 \Sigma(t,t'')dt'' \label{Wo}\ee so that \be \Sigma(t,t') = \frac{d}{dt'}W_0(t,t'),\quad W_0(t,0)=0. \label{rela} \ee Integrating by parts in eq.(\ref{intdiff}) we obtain \be \int_0^t \Sigma(t,t')\,C_A(t')\, dt' = W_0(t,t)\,C_A(t) - \int_0^t W_0(t,t')\, \frac{d}{dt'}C_A(t') \,dt'. \label{marko1}\ee The second term on the right hand side is formally of \emph{fourth order} in $H_I$ suggesting how  a systematic approximation scheme can be developed. Setting \be W_1(t,t') = \int^{t'}_0 W_0(t,t'') dt'', \Rightarrow \frac{d}{dt'} W_1(t,t')= W_0(t,t');  \quad W_1(t,0) =0 \,\label{marko2} \ee and integrating by parts again, we find \be \int_0^t W_0(t,t')\, \frac{d}{dt'}C_A(t') \,dt' = W_1(t,t)\,\dot{C}_A(t) +\cdots \label{marko3} \ee leading to   \be \int_0^t \Sigma(t,t')\,C_A(t')\, dt' = W_0(t,t)\,C_A(t) - W_1(t,t)\,\dot{C}_A(t) +\cdots \label{histoint}\ee

This process can be implemented systematically resulting in higher order differential equations. Since $W_1 \simeq H^2_I \,;\, \dot{C}_A \simeq H^2_I$ the second term in (\ref{histoint}) is $\simeq H^4_I$. We consistently neglect this term because to this order the states $\ket{\kappa}$ may also have non-vanishing matrix elements with states $\ket{\kappa'}$ other than $\ket{A}$ and the hierarchy would have to include these other states, therefore yielding contributions of $\mathcal{O}(H^4_I)$.  Hence up to leading order $\simeq H^2_I$   the equation eq.(\ref{intdiff}) becomes \be \dot{C}_A(t)= -  W_0(t,t) C_A(t)   \label{markovian}\ee   where
\be W_0(t,t) = \int^{\infty}_{-\infty} \rho(p_0) \Bigg[\frac{1-e^{-i(p_0-E_A)t}}{i(p_0-E_A)} \Bigg]\,dp_0\,,\label{Wzerot}\ee yielding
\be C_A(t) = e^{-it\,\delta E(t) }\, e^{-\frac{\gamma(t)}{2}} \,, \label{solumarkov}\ee where we used that $C_A(0)=1$,  with
\be \delta E(t) = \int^{\infty}_{-\infty} \frac{\rho(p_0)}{(E_A-p_0)}\,\Bigg[1- \frac{\sin\Big(\big(E_A-p_0 \big)t\Big) }{(E_A-p_0) \big)t} \Bigg]\, dp_0\,, \label{real} \ee and
\be \gamma(t) = 2\,\int^{\infty}_{-\infty} \rho(p_0) \frac{\Big[1-\cos\Big(\big(E_A-p_0 \big)t\Big) \Big]}{\big(E_A-p_0\big)^2}\,dp_0 \,. \label{imag} \ee The survival probatility of the initial state is given by
\be |\langle{A}|{\Psi(t)}\rangle|^2 = |C_A(t)|^2 = e^{-\gamma(t)}\,. \label{surviprob}\ee

In the long time limit

\be \delta E(t)_{~~ \overrightarrow{t\longrightarrow \infty}~~} \,\delta E_{\infty} = \int^{\infty}_{-\infty} \mathcal{P}\, \frac{\rho(p_0)}{(E_A-p_0)}\,dp_0 \,,\label{asyreal}\ee  where $\mathcal{P}$ stands for the principal part,  yielding a renormalization of the bare frequency of the state $A$, namely $E_A+ \delta E_{\infty} = E_{AR}$, whereas the long time limit of $\gamma(t)$   yields the decay law of the initial state.

 The spectral density is only non-vanishing for $p_0 \geq E_T$ where $E_T$ is the beginning of the multiparticle threshold. The long time limit of (\ref{imag}) is dominated by the region of the spectral density $p_0 \simeq E_A$, therefore it depends on whether $E_A < E_T$ or $E_A \geq E_T$.

 \vspace{1mm}

 \textbf{i) $E_A < E_T$:} in this case      the oscillatory function averages out in the long time limit since the region $p_0\simeq E_A$ is not within the region of support of the spectral density, therefore

\be \gamma(t)_{~~ \overrightarrow{t\longrightarrow \infty}~~} \, z_A = 2 \int^{\infty}_{E_T}  \, \frac{\rho(p_0)}{(E_A-p_0)^2}\,dp_0 \,,\label{zA}\ee yielding
\be |C_A(t)|^2_{~~ \overrightarrow{t\longrightarrow \infty}~~} \mathcal{Z}_A = e^{-z_A} \,,\label{wavefun1} \ee where $\mathcal{Z}_A$ is the wave function renormalization. Since $\rho(p_0) \geq 0$ (see eqn. (\ref{rhopo})) $z_A >0$ and $\mathcal{Z}_A < 1$ consistently with the unitarity condition (\ref{unitarity1}). This case describes a stable particle, with its mass shell described by an isolated pole below the multiparticle threshold.

\vspace{1mm}

 \textbf{ii) $E_A > E_T$:}  in this case,  $\rho(E_A) \neq 0$, and  the long time limit is dominated by the neighborhood  of $E_A$, subtracting $\rho(E_A)$ from the spectral density,  we find in the long time limit
 \be \gamma(t)_{~~ \overrightarrow{t\longrightarrow \infty}~~} \Gamma_A\,t +  z_A + \mathcal{O}(1/t) + \cdots \,,\label{GA}\ee where
\be  \Gamma_A = 2\pi \rho(E_A)~~;~~ z_A = 2 \int^{\infty}_{-\infty}  \, \mathcal{P}\,\frac{\rho(p_0)}{(E_A-p_0)^2}\,dp_0 \,,\label{Gaz} \ee yielding
\be |C_A(t)|^2_{~~ \overrightarrow{t\longrightarrow \infty}~~} \mathcal{Z}_A \,e^{-\Gamma_A t} ~~;~ \mathcal{Z}_A = e^{-z_A} \,.\label{decay} \ee Therefore this case describes an unstable, decaying state, namely a resonance.

\vspace{1mm}

 \textbf{iii) $E_A = E_T$:} in this case the multiparticle threshold coincides with the position of the mass and the spectral density vanishes at $p_0 = E_A$. The long time dynamics is now determined by how the spectral density vanishes at threshold. In the case that the spectral density vanishes \emph{linearly} at threshold, the $p_0$ integral in $\gamma(t)$ (eqn. (\ref{imag})) features a logarithmic divergence at long time. This is the case for the super-renormalizable    and renormalizable cases (eqns. (\ref{Gnearpole},\ref{nearpole}) respectively)   studied in the previous section, where the discontinuity of the propagator across the two-particle cut vanishes linearly  at threshold in the variably $p_0-E_P$, namely
  \be \rho(p_0)_{~~ \overrightarrow{p_0\longrightarrow E_A}~~} \Delta (p_0-E_A) ~~;~~ \Delta = \Big[d \rho(p_0)/dp_0\Big]_{p_0=E_A}\,.  \label{linthres}\ee

     To understand this case more clearly,   and to extract the infrared divergence,  it proves convenient to change variables    to $(p_0-E_A) = s\,E_A$ with $\rho(p_0) \equiv \overline{\rho}(s)$  and $\tau = E_A\,t$, yielding
 \be \gamma(t) = \frac{2}{E_A} \,  \int^{\infty}_0 \overline{\rho}(s)\,\frac{1-\cos(s \tau)}{s^2}\,ds = \frac{2}{E_A} \,  \int^{1}_0 \overline{\rho}(s)\,\frac{1-\cos(s \tau)}{s^2}\,ds + \frac{2}{E_A} \,  \int^{\infty}_1 \overline{\rho}(s)\,\frac{1-\cos(s \tau)}{s^2}\,ds\,.  \label{gamsvars} \ee The first integral features an infrared divergence, whereas the second is infrared finite and the cosine term averages out in the long time limit. With the threshold behavior (\ref{linthres}), let us write for the first integral
 \be \overline{\rho}(s) = \Delta\, E_A\,s + \widetilde{\rho}(s)~~;~~ \widetilde{\rho}(s)_{~~ \overrightarrow{s\longrightarrow 0}~~} \propto s^{n}~~;~~ n\geq 2 \,,\label{separo}\ee
 leading to
 \be \gamma(t) = 2 \,\Delta \int^{1}_0 \frac{1-\cos(s \tau)}{s}\,ds + \mathcal{F}(\tau)    _{~~ \overrightarrow{t\longrightarrow \infty}~~} 2\Delta \,\ln\big[E_A\,t\big] +  {z}_A\,, \label{anomalous} \ee where the remaining function $\mathcal{F}(\tau)_{~~ \overrightarrow{\tau \longrightarrow \infty}~~}\mathcal{F}_{\infty}$   a time-independent asymptotic long time limit. This case leads to the relaxation of the amplitude with an \emph{anomalous dimension}, namely
 \be |C_A(t)|^2_{~~ \overrightarrow{t\longrightarrow \infty}~~} \Big[E_A t\Big]^{-2\Delta}\,  {\mathcal{Z}}_A~~;~~  {\mathcal{Z}}_A = e^{-{z}_A} \,,\label{anomdim} \ee in agreement with the results (\ref{ltG},\ref{Soft1posE}) obtained by the inverse Fourier transform of a \emph{renormalization group improved} propagator. Therefore this dynamical resummation method provides a real time implementation of the renormalization group. The wave function renormalization constant ${\mathcal{Z}}_A~$ is infrared finite, however it is ultraviolet divergent in a renormalizable (or non-renormalizable) theory.

 If $\Delta =0$ and the spectral density vanishes faster than linear near threshold it follows from the above result that $\gamma(t) \rightarrow z_A$ in the long time limit.

 \subsection{Equivalence with the dynamical renormalization group}\label{subsec:drg}

 In the interaction picture, the time evolution of a state is given by
 \be |\psi(t)\rangle = U(t,t_0) \; |\psi(t_0)\rangle \,, \label{timeevi} \ee  where
 \be U(t,t_0)= 1 -i  \;  \int^t_{t_0} {H}_I(t') \; dt'-
\; \int^t_{t_0}\int^{t'}_{t_0} {H}_I(t') \;  {H}_I(t'') \; dt' \; dt'' +
{\cal O}(H^3_I) ~~,~~ U(t_0,t_0)=1 \; . \label{unitop}\ee

If at $t_0=0$ the initial state is $\ket{\psi(t_0)} = C_A(0) \ket{A}$ the survival amplitude at time $t$ is given by
\bea && C_A(t) =  C_A(0) \langle A|U(t,0) \ket{A}  =
C_A(0)\Bigg[1 - i\;
     t \;  \langle A|H_I(0)|A \rangle
  \nonumber \\&& - \; \int^t_{0} \int^{t'}_{0}\sum_{\kappa\ } |\langle A|H_I(0)|\kappa\rangle|^2 \;  e^{i(E_A-E_\kappa)(t'-t'')} \; dt'
\; dt'' + {\cal O}(H^3_I)\Bigg]  \; ,\eea where we have introduced $\sum_{\kappa} \ket{\kappa}\bra{\kappa} = 1$ in the second order term, and introduced the initial amplitude $C_A(0)$ to clarify the nature of the dynamical renormalization group. We will be mainly concerned with the examples discussed in the previous sections, for which $\bra{A}H_I(0)\ket{A}=0$. In terms of the spectral density (\ref{rhopo}) we find
\be  C_A(t) =  C_A(0)\Big\{1   + \mathcal{S}^{(2)}_A(t) +{\cal O}(H^3_I)
 \Big\}\;,\label{CAoftdrg}
\ee  where
\be  \mathcal{S}^{(2)}_A(t)   =    -i  \, t \; \delta E(t)-\frac{1}{2}\,\gamma(t) \,,  \label{secular}\ee  where the superscript in $\mathcal{S}$ refers to the order in perturbation theory and $\delta E(t);\gamma(t)$ are given by eqns. (\ref{real},\ref{imag}) respectively. From the results (\ref{asyreal},\ref{GA},\ref{anomalous}) it follows that $\mathcal{S}^{(2)}_A(t)$ features \emph{secular growth in time},  namely  it grows in time invalidating the reliability of the perturbative expansion at long time. The dynamical renormalization group\cite{gold,goldbook,boyvega} provides a resummation framework that improves the convergence at long time. It begins by evolving in time  up to a time $\tau$ long enough to establish the secular growth, but short enough so that perturbation theory is still reliable and absorbing the time evolution into a renormalization of the amplitude. This program is implemented as follows, writing
  \begin{equation}
C_A(0)= C_A(\tau) \; R_A(\tau)~~, ~~ R_A(\tau)= 1+
   r_A^{(2)}(\tau)+{\cal O}(H^3_I)
\; , \label{renoamp}
\end{equation} where $r_A^{(2)}(\tau) \simeq \mathcal{O}(H^2_I)$, etc. Up to second order in the interaction we obtain
\be C_A(t) = C_A(\tau) \Big[1 + \big(r_A^{(2)}(\tau) + \mathcal{S}^{(2)}_A(t)  \big) + \cdots \Big]\,, \label{Cren} \ee the counterterm $ r_A^{(2)}(\tau)$ is  chosen to cancel   $\mathcal{S}^{(2)}_A(t)$ at the renormalization time scale $t=\tau$, namely
\be r_A^{(2)}(\tau) =-  \mathcal{S}^{(2)}_A(\tau)\,. \label{renscheme}\ee The time dependent amplitude $C_A(t)$ \emph{does not} depend on the arbitrary renormalization scale $\tau$, hence
\be \frac{d}{d\tau} C_A(t) =0 \,, \label{drg} \ee this is the dynamical renormalization group equation. Consistently keeping up to terms of $\mathcal{O}(H^2_I)$ this equation  leads to
\be \frac{\dot{C}_{A}(\tau)}{C_A(\tau)}= -  \dot{r}_A^{(2)}(\tau) + \mathcal{O}(H^4_I) + \cdots \label{drg2}\ee where the dots stand for $d/d\tau$. Using the renormalization condition (\ref{renscheme}) the solution is given by
\be C_A(\tau) = C_A(\tau_0)~e^{\big(\mathcal{S}^{(2)}_A(\tau)-\mathcal{S}^{(2)}_A(\tau_0)\big)}\,, \label{soldrg}\ee now we can choose $\tau = t~;~ \tau_0 =0$ and $ C_A(0) =1$
with the result
\be C_A(t) = e^{-it\,\delta E(t) }\, e^{-\frac{\gamma(t)}{2}}\,,\label{drgsolfin}\ee which is precisely the result given by eqn. (\ref{solumarkov}). This solution provides a resummation of the perturbative series up to second order in the coupling. In the case in which the mass shell is embedded in the particle continuum, the long time behavior of the amplitude is $C_A(t) = \mathcal{Z}_A\, e^{-i\Delta E_\infty t}\,e^{-\Gamma t/2}$ yielding the usual exponential  decay law in agreement with eqn. (\ref{decay}). Therefore, the   dynamical resummation method described in the previous section is equivalent to the dynamical renormalization group resummation of secular terms. However, a bonus of the dynamical resummation method is that it also yields the coefficients $C_{\kappa}(t)$ given by eqn. (\ref{Ckapas}), and a  direct connection with unitarity (see eqn. (\ref{unitarity1})).

 Armed with these general results, we now address the cases studied in the previous section.

 \subsection{Super-renormalizable case:}\label{subsec:supren}  The interaction Hamiltonian in the interaction picture for the model described by eqn. (\ref{lagsuper}) is
\be  H_I(t) = \lambda\,\int d^3 x \,\phi^\dagger(\vx,t)\,\phi(\vx,t)\,\chi(\vx,t) \,,\label{HIsuper} \ee where the time evolution is that of free fields. In this case the state $\ket{A} = \ket{1^{\phi}_{\vp}}$ i.e. a single particle state of the field $\phi$ and the states $\ket{\kappa} = \ket{1^{\phi}_{\vk}\,;\,1^{\chi}_{\vq}}$ ,  a two particle intermediate state. We quantize the fields in a volume $V$ with a discrete momentum representation,  eventually $V$ is taken to infinity.  The matrix element
\be \bra{1^{\phi}_{\vp}} H_I(0)\ket{1^{\phi}_{\vk}\,;\,1^{\chi}_{\vq}} = \frac{\lambda\,V\,\delta_{\vp,\vk+\vq}}{\Big[2V E_p 2 V E_k 2V \omega_q \Big]^{1/2}} \,, \label{mtxel1}\ee where $E_p=\sqrt{p^2+M^2},\omega_q= |\vq| $ are the energies of the $\phi, \chi$ particle respectively, and the total energy of this intermediate state is $E_{\kappa} = E_k+\omega_q$. With $\sum_{\kappa} = \sum_{\vp}\sum_{\vq}$ the spectral density (\ref{rhopo}) is given by
\be \rho(p_0;p) = \frac{\lambda^2}{8E_p} \,\int \frac{d^3k}{(2\pi)^3}\,\frac{\delta(p_0-E_k-|\vp-\vk|)}{E_k\,|\vp-\vk|}\,. \label{rhosup}\ee For $p_0 = E_p$ this is identified with the Lorentz invariant phase space for two body decay, which must vanish by kinematics because a massive particle cannot emit or absorb a massless particle on shell. Therefore the spectral density must vanish as $  p_0\longrightarrow E_p $.

We find
\be \rho(p_0;p) = \frac{\lambda^2}{32\,\pi^2\,E_p}\,(p_0-E_p) \Bigg( \frac{p_0+E_p}{p^2_0-p^2}\Bigg)\,\Theta(p_0-E_p) \,,\label{rhosupfin}\ee vanishing linearly as $\Delta \, (p_0-E_p)$ at threshold with $\Delta = (\lambda/4\pi M)^2$. Introducing the variables
$s =  (p_0-E_p)/E_p ~~;~~   T = E_p\,t  ~~;~~ R=M/E_p$  we find that the function $\gamma(t)$ in eqn. (\ref{imag}) can be written as
\be \gamma(t) = I_1(T)+ I_2(T) \,,\label{Ifuncs}\ee with
\bea I_1(T) & = & \Delta \, R^2\, \int^1_0\,\Big[ \frac{2+s}{R^2+2s+s^2}\Big]\Big[ \frac{1-\cos(s\,T)}{s}\Big]\, ds \label{Iuno} \\
I_2(T) & = & \Delta \, R^2\,\int^{\infty}_1\,\Big[ \frac{2+s}{R^2+2s+s^2}\Big]\Big[ \frac{1-\cos(s\,T)}{s}\Big]\, ds \,. \label{Idos} \eea The integral $I_1(T)$ features an infrared divergence at $s =0$, which  can be isolated by subtracting the first bracket inside the integral in (\ref{Iuno}) at $s=0$, yielding
\be I_1(T)  =  2 \Delta  \int^1_0 \Big[\frac{1-\cos(s\,T)}{s}\Big]\, ds +  \Delta \int^1_0 \Big[ \frac{R^2-4-2s}{R^2+2s+s^2}\Big]\,(1-\cos(s\,T)) \, ds \,.\label{I1sepa}\ee In the long time limit the $\cos(sT)$ terms in $I_2$ and the second term in (\ref{I1sepa}) average out, yielding
\be   \gamma(t)_{~~ \overrightarrow{t\longrightarrow \infty}~~}  2 \,\Delta\, \ln\big[E_p t \big]+ z_{\phi}\,, \label{gamafilt} \ee where
\be z_{\phi} = \Delta\, \Bigg\{2\gamma_E + \int^1_0 \Big[ \frac{R^2-4-2s}{R^2+2s+s^2}\Big] \, ds + R^2\,  \int^{\infty}_1\,\Big[ \frac{2+s}{R^2+2s+s^2}\Big] \, \frac{ds}{s}  \Bigg\}\,, \label{zfi} \ee with $\gamma_E$ the Euler-Mascheroni constant. Therefore the long time behavior of the survival probability is given by
\be |C_{\phi}(t)|^2 = \Big[E_p\,t \Big]^{-2\Delta}\,\mathcal{Z}_{\phi}~~;~~\mathcal{Z}_{\phi}= e^{-z_{\phi}}\,, \label{Cfi2}\ee displaying the power law decay of the probability with anomalous dimension $2\,\Delta$ in complete agreement with the result from the renormalization group improved propagator eqn. (\ref{proba1}). We note that the wave function renormalization $\mathcal{Z}_{\phi}$ is infrared finite and also  ultraviolet finite  as befits a super renormalizable theory.

 \subsection{Renormalizable case:} For the renormalizable  case described by the Lagrangian density (\ref{yukaL}) the interaction Hamiltonian in the interaction picture of free fields is
 \be H_I(t) = Y\,\int d^3 x \overline{\Psi}(\vx,t) \Phi(\vx,t) \Psi(\vx,t)\,, \label{HIren}\ee with the state $\ket{A} = \ket{1^{\psi}_{\vp,\alpha}}$ and the intermediate states $\ket{\kappa} = \ket{1^{\psi}_{\vk,\beta};1^{\phi}_{\vq}}$. The matrix elements are given by
 \bea \bra{1^{\psi}_{\vk,\beta};1^{\phi}_{\vq}}H_I(0)\ket{1^{\psi}_{\vp,\alpha}} & = & V\, Y \delta_{\vp,\vk+\vq}~~
 \frac{\overline{\mathcal{U}}_{\vk,\beta,a}\,{\mathcal{U}}_{\vp,\alpha,a}}{\Big[ 2VE_p 2V E_k 2V |\vq|\Big]^{1/2}}\, \label{mtxelren1}\\
 \bra{1^{\psi}_{\vp,\alpha}}H_I(0)\ket{1^{\psi}_{\vk,\beta};1^{\phi}_{\vq}} & = & V\, Y \delta_{\vp,\vk+\vq}~~\frac{\overline{\mathcal{U}}_{\vp,\alpha,b}\,{\mathcal{U}}_{\vk,\beta,b}}{\Big[ 2VE_p 2V E_k 2V |\vq|\Big]^{1/2}}\,. \label{mtxelren2AX} \eea With $\sum_{\kappa} = \sum_{\vk}\sum_{\vq}\sum_{\beta}$ and averaging over the initial polarizations $\alpha$, the spectral density (\ref{rhopo}) becomes
 \be \rho(p_0;p) = \frac{Y^2}{4E_p} \, \int \frac{d^3k}{(2\pi)^3}\,\frac{\delta(p_0-E_k-|\vp-\vk|)}{E_k\,|\vp-\vk|}\,\Big[k\cdot p+M^2 \Big]\,,  \label{rhoren}\ee which is found to be
 \be \rho(p_0;p)   =    \frac{Y^2}{32\pi^2 E_p}\,(p_0-E_p) \, \Big[\frac{p_0+E_p}{p^2_0-p^2}\Big]\,\Bigg\{p_0\,
 \Big[\frac{E_p-p_0}{p^2_0-p^2}\Big](p^2_0-p^2 +M^2)+p^2_0-p^2+3M^2 \Bigg\}   \Theta(p_0-E_p)   \,. \label{rhofinren}\ee  We note that for $p_0 \simeq E_P$, $\rho(p_0,p) \simeq \Delta  (p_0-E_p) + \cdots$  where $\Delta = Y^2/(4\pi^2) $ and the dots stand for terms that vanish as $(p_0-E_p)^{n}\,,\,n\geq 2$ near threshold. To separate the infrared contribution, we change variables to $p_0-E_p = s E_p ~~;~~ \rho(p_0) \equiv \overline{\rho}(s)$  and  $T = E_p\,t$ and
 write
 \be \overline{\rho}(s) = \Delta\, E_p \, s    + \widetilde{\rho}(s)~~;~~ \Delta = \frac{Y^2}{4\pi^2}\,, \label{IRren} \ee with $\widetilde{\rho}(s)_{~~ \overrightarrow{s\longrightarrow 0}~~} \propto s^n ~~;~~ n\geq 2$ , yielding
 \be \gamma(t ) = J_1(T) + J_2(T) \label{gamsplit}\ee with
 \be J_1(T) =  2\,\Delta \, \int^1_0 \frac{1-\cos(sT)}{s}\,ds = 2\,\Delta \Big\{ \ln[E_p t] +\gamma_E - Ci[E_p t] \Big\} \,,\label{Juno} \ee
 \be J_2(T) =  2 \int^1_0 \widetilde{\rho}(s)\, \frac{1-\cos(sT)}{s^2}\,ds + 2 \int^\infty_1  \overline{\rho}(s)\, \frac{1-\cos(sT)}{s^2} \,ds\,.  \label{Jdos}\ee

 In $J_2(T)$ the cosine term averages out in the long time limit and this conribution approaches a time independent asymptotic value, which however is ultraviolet divergent because $\overline{\rho}(s) \simeq s$ as $s\rightarrow \infty$. Therefore in the long time limit we find
 \be \gamma(t) _{~~ \overrightarrow{t\longrightarrow \infty}~~} 2\Delta \,\ln\big[E_p\,t\big] + \widetilde{z}_\psi   \,,  \ee yielding the survival probability
 \be |C_{\psi}(t)|^2 = \mathcal{Z}_{\psi}\, \big[E_p\,t\big]^{-2\,\Delta} \label{probafer}\ee which agrees with the power law decay of the amplitude with anomalous dimension  in eqn. (\ref{Soft1posE}).  In this case the wave function renormalization $\mathcal{Z}_{\psi}$ is infrared finite but ultraviolet divergent since this is a renormalizable theory.

\subsection{Axion couplings:} In the case of the pseudoscalar coupling $ig \bar{\Psi}\gamma^5\phi\Psi$ the matrix elements are given by
 \bea \bra{1^{\psi}_{\vk,\beta};1^{\phi}_{\vq}}H_I(0)\ket{1^{\psi}_{\vp,\alpha}} & = & ig\, V\, \delta_{\vp,\vk+\vq}~~
 \frac{\overline{\mathcal{U}}_{\vk,\beta,a}\,\gamma^5 \, {\mathcal{U}}_{\vp,\alpha,a}}{\Big[ 2VE_p 2V E_k 2V |\vq|\Big]^{1/2}}\, \label{mtxelren1AX}\\
 \bra{1^{\psi}_{\vp,\alpha}}H_I(0)\ket{1^{\psi}_{\vk,\beta};1^{\phi}_{\vq}} & = & ig\, V\,  \delta_{\vp,\vk+\vq}~~\frac{\overline{\mathcal{U}}_{\vp,\alpha,b}\,\gamma^5\,{\mathcal{U}}_{\vk,\beta,b}}{\Big[ 2VE_p 2V E_k 2V |\vq|\Big]^{1/2}}\,. \label{mtxelren2} \eea With $\sum_{\kappa} = \sum_{\vk}\sum_{\vq}\sum_{\beta}$ and averaging over the initial polarizations $\alpha$, the spectral density (\ref{rhopo})  now becomes
 \be \rho(p_0;p) = \frac{g^2}{4E_p} \, \int \frac{d^3k}{(2\pi)^3}\,\frac{\delta(p_0-E_k-|\vp-\vk|)}{E_k\,|\vp-\vk|}\,\big[k\cdot p-M^2 \big]\,.  \label{rhorenAX}\ee Following the same steps as in the scalar Yukawa coupling case we find
 \be  \rho(p_0;p)   =    \frac{g^2}{32\pi^2 E_p}\,(p_0-E_p)^2 \, \Big[\frac{p_0+E_p}{p^2_0-p^2}\Big]\Bigg\{p_0 + E_p - p_0 \Big[\frac{p^2_0-p^2 +M^2}{p^2_0-p^2} \Big]  \Bigg\} \Theta(p_0 -E_p) \,. \label{noirps} \ee

In this case the $(p_0-E_p)^2 $ completely cancels the denominator in $\gamma(t)$ eqn. (\ref{imag}) therefore there are no infrared singularities and the asymptotic long time limit $\gamma(t) \rightarrow z_\psi$ which, however, is ultraviolet divergent. The behavior $\propto (p_0 - E_P)^2$ near threshold is, as expected,  in complete agreement with the result (\ref{imcut1}) of the self energy.

For the pseudovector coupling $g\partial_{\mu}\phi \bar{\Psi}\gamma^\mu\gamma^5 \Psi$ we now find
\be \rho(p_0;p) = \frac{g^2}{4E_p} \, \int \frac{d^3k}{(2\pi)^3}\,\frac{\delta(p_0-E_k-|\vp-\vk|)}{E_k\,|\vp-\vk|}\,\big[(k\cdot q)\,(p\cdot q) \big]~~;~~q^\mu = (|\vp-\vk|;\vp-\vk)\,,  \label{rhorenAXvec}\ee with the result
\be \rho(p_0,p) = \frac{g^2}{32\pi^2}\, (p_0-E_p)^3\, \Big[\frac{(p_0+E_p)^2}{p^2_0-p^2} \Big]\,\Big[1 + \frac{M^2 }{p^2_0-p^2} \Big]\,\Theta(p_0-E_p)\,. \label{veraxth} \ee
The behavior $\simeq (p_0 -E_p)^3$ near threshold   is consistent with the results (\ref{dercoupir},\ref{imcut2}), and  implies that  in this case there is no infrared singularity, and furthermore $  \gamma(t)_{~~ \overrightarrow{t\longrightarrow \infty}~~} z_{\psi} $ with $z_\psi$ being ultraviolet divergent.   Therefore, we conclude that either  pseudoscalar or pseudovector  axion couplings do not yield infrared divergences.

\vspace{1mm}

\textbf{Criterion for infrared divergences:} The study of the previous sections allows us to provide a general criterion to determine which type of (non-gauge) interactions even from effective field theories yield infrared divergences at one loop level and which ones do not. The typical form of the spectral density at one loop level can be written as
\be \rho(p_0) \propto \int \frac{d^3k}{(2\pi)^3}\,\frac{\delta(p_0-E_k-\omega_q)}{E_k\,\omega_q}\,
\mathcal{F}[(k\cdot p);(k\cdot q);(q\cdot p)] \,,\label{generalrho}\ee where $\mathcal{F}$ is a Lorentz invariant function of the scalar products of the on-shell four vectors $p^\mu = (E_p,\vp)~;~k^\mu= (E_k,\vk)~~;~~ q^\mu =    (\omega_q;\vp-\vk)~;~\omega_q=|\vp-\vk|$. The most general form of $ \mathcal{F}[(k\cdot p);(k\cdot q);(q\cdot p)]$ is a combination of polynomials, namely
\be \mathcal{F}[(k\cdot p);(k\cdot q);(q\cdot p)] = \sum_{m.n,l} a_{mnl} \,(k\cdot p)^m\,(k\cdot q)^m \,(q\cdot p)^l ~~;~~ (m,n,l) = 0,1,2\cdots  \,, \label{geneF}\ee with $a_{mnl} $ coefficients that depend on the particular effective field theory interaction. As it will become clear below,  it suffices to consider only monomials in these products.

The angular integration in (\ref{generalrho}) is performed yielding
\be \rho(p_0) \propto \frac{1}{4\pi^2\,p} \int^{E_+}_{E_-} \, \mathcal{F}[(k\cdot p);(k\cdot q);(q\cdot p)]\big|_{\omega_q=p_0-E_k}\,dE_k\,, \label{psint}\ee where
\be E_{\pm} = \frac{(p_0\pm p)^2+M^2}{2(p_0\pm p)} \Rightarrow E_+ - E_- = (p_0-E_p)\, \frac{p\,(p_0+E_p)}{p^2_0-p^2} \,. \label{Epm}\ee

 The vanishing of the integral as $p_0 \rightarrow E_p$ reflects the vanishing of the phase space for on-shell emission of the massless quanta. As discussed above, the infrared divergence arises from the contribution to the spectral density that vanishes linearly at threshold. Since $E_+-E_-$ vanishes linearly as $p_0 \rightarrow E_p$, only terms in $\mathcal{F}[(k\cdot p);(k\cdot q);(q\cdot p)]$ that remain finite in this limit yield infrared divergences. Therefore it is now a matter of analyzing the behavior of the various scalar products to reveal which contributions do yield infrared divergences. We find (using the delta function constraint in (\ref{psint}))
\bea p\cdot k & = & \frac{1}{2}(p_0-E_p)\big[p_0+E_p-2E_k \big]+M^2_{ ~~\overrightarrow{{p_0 \rightarrow E_p}}}~~  M^2 ~~\Rightarrow~~ \mathrm{IR}~\mathrm{div}   \label{irdivkp} \\
 p\cdot q & = & \frac{1}{2}(p_0-E_p)\big[-p_0+E_p+2E_k \big]_{ ~~\overrightarrow{{p_0 \rightarrow E_p}}}  ~~\propto (p_0-E_p) ~~\Rightarrow~~ \mathrm{NO}~\mathrm{IR}~\mathrm{div} \label{noirdivpq}\\  k\cdot q & = & \frac{1}{2}(p_0-E_p)\big[p_0+E_p \big]_{ ~~\overrightarrow{{p_0 \rightarrow E_p}}} ~~\propto (p_0-E_p) ~~\Rightarrow~~ \mathrm{NO}~\mathrm{IR}~\mathrm{div} \,.  \label{noirdivkq} \eea

 This analysis explains why the scalar Yukawa coupling with $\mathcal{F} = p\cdot k+M^2$ yields an infrared divergence whereas the pseudoscalar axion Yukawa coupling with $\mathcal{F} = p\cdot k-M^2$ does not. It also reveals that effective field theories with derivative couplings that necessarily yield $\mathcal{F} \propto p\cdot q~;~k \cdot q$ do not yield infrared divergences at one loop order. It is important to highlight that these arguments are only valid at one loop level in \emph{non-gauge} theories, we were not able to extend them generically beyond this order in perturbation theory.

\section{Unitarity and dressing cloud}\label{sec:unitarity} In the cases in which the infrared divergences at threshold lead to the decay  of the single particle amplitude with a power law with anomalous dimension, at long time this
amplitude vanishes. Unitarity must be fulfilled by ``populating'' the intermediate states with amplitudes $C_{\kappa}(\infty)$  such that $\sum_{\kappa} |C_{\kappa}(\infty)|^2 = 1$. The fulfillment of unitarity when the amplitude of the initial state vanishes altogether, and the coefficients $|C_{\kappa}|^2$ being formally of $\mathcal{O}(\Delta)$ implies that the sum over the intermediate states must be proportional to $1/\Delta$. This integral must be singular in the limit $\Delta \rightarrow 0$ bringing about a non-perturbative cancellation of the $\Delta$ from the coefficients. We now study how this result from unitarity emerges in the long time limit.

 From eqn. (\ref{Ckapasol}) we find
\be C_{\kappa}(t) = -i \bra{\kappa}H_I(0)\ket{A} \int^t_0 e^{ i\Omega t'}\,C_A(t')\,dt' ~~;~~ \Omega = E_\kappa-E_A \,,\label{ckap2}\ee and
\be |C_{\kappa}(t)|^2 = |\bra{A}H_I(0)\ket{\kappa}|^2 \int^t_0 \int^t_0  e^{i\Omega t_1} \, C_A(t_1)\,e^{-i\Omega t_2} \, C^*_A(t_2)\,dt_1\,dt_2 \,. \label{modC2}\ee Inside the integrals we replace the amplitudes $C_A(t)$ by eqn. (\ref{solumarkov}). Since at early time  the amplitude departs from   $C_A(0) =1$ by a perturbatively small amount, we will replace them by the long time limit (\ref{anomalous})  \be C_A(t) = e^{-i\delta E_{\infty}\,t}\, e^{-\frac{\gamma(t)}{2}} ~~;~~  \gamma(t) = 2\Delta \ln\big[E_A t\big] + z_A  \,,\label{ampslt}\ee (see eqn. (\ref{anomdim})) and absorb $\delta E_{\infty}$  into a renormalization of $E_A$ (mass renormalization). The integrand in the double time integral in  (\ref{modC2}) is now given by ($E_A$ in $\Omega$ now stands for the renormalized energy)
\be J(t_1,t_2) = e^{i\,\Omega(t_1-t_2)}\,e^{-\frac{1}{2}(\gamma(t_1)+\gamma(t_2))}\,, \label{J12}\ee writing the double time integral in (\ref{modC2}) as
\be \int^t_0 \int^t_0 J(t_1,t_2) \,\Big(\Theta(t_1-t_2)+\Theta(t_2-t_1)\Big)\,dt_1\,dt_2 = 2 \int^t_0 dt_1 e^{-\frac{\gamma(t_1)}{2}}\int^{t_1}_0 \cos[\Omega(t_1-t_2)]\,e^{-\frac{\gamma(t_2)}{2}} \,   dt_2 \,,\label{dfun}\ee where in the term with $\Theta(t_2-t_1)$ on the left hand side of (\ref{dfun}) we relabelled $t_1 \leftrightarrow t_2$ and used that $J(t_2,t_1) = J^*(t_1,t_2)$ with $\gamma(t)$ being real. Now writing
\be \cos[\Omega(t_1-t_2)] = \frac{d}{dt_2} G[t_1;t_2] ~~;~~ G[t_1;t_2] = \int^{t_2}_0 \cos[\Omega(t_1-t')]\,dt' ~~;~~ G[t_1;0]=0 \,,\label{Gf}\ee in the $t_2$ integral in (\ref{dfun}), integrating by parts using (\ref{Gf}) and neglecting the term proportional to the time derivative of $\gamma(t_2)$ because it is
of $\mathcal{O}(H^2_I)$, hence consistently neglecting terms of $\mathcal{O}(H^4_I)$ in (\ref{modC2}) we find that the double integral in (\ref{modC2}) becomes
\be \int^t_0 \int^t_0  e^{-i\Omega t_1} \, C_A(t_1)\,e^{i\Omega t_2} \, C^*_A(t_2)\,dt_1\,dt_2 = \frac{2\,\mathcal{Z}_A}{\Omega} \,\int^t_0 \sin\big[\Omega\,t_1\big]\,\big[E_At_1\big]^{-2\Delta}\, dt_1\,. \label{findoubint}\ee In the limit $t\rightarrow \infty$ in (\ref{findoubint}) we can rescale $\Omega t_1 \equiv u$ and using a representation of the gamma function we find
\be  |C_{\kappa}(\infty)|^2 = 2 \, \mathcal{Z}_A\, \Bigg[ \frac{|\bra{A}H_I(0)\ket{\kappa}|^2}{(E_\kappa-E_A)^2}\Bigg]\, \Big[\frac{E_{\kappa}-E_A}{E_A} \Big]^{2\Delta}\, \Gamma[1-2\Delta]\,\sin\big[\frac{\pi}{2}(1-2\Delta) \big]\,. \label{Ckapinfty}  \ee In terms of the spectral density (\ref{rhopo}) and introducing $\overline{Z}_A =\mathcal{Z}_A\,\Gamma[1-2\Delta]\,\sin\big[\frac{\pi}{2}(1-2\Delta) \big]$ we obtain
\be \sum_{\kappa}|C_{\kappa}(\infty)|^2 = 2\,\overline{Z}_A\,\int^{\infty}_{-\infty} \frac{\rho(p_0)}{(p_0-E_A)^2}\,\Big[\frac{p_0-E_A}{E_A}\Big]^{2\Delta} \, dp_0 \,,  \ee This is the general result for the cases with infrared divergences at threshold. We now apply this result to the super renormalizable case as an example,  of which the renormalizable case is a simple extension. In this case (see section (\ref{subsec:supren})) the state  $\ket{A} = \ket{1^{\phi}_{\vp}}$ and the states $\ket{\kappa} = \ket{1^{\phi}_{\vk}\,;\,1^{\chi}_{\vq}}$ with energy $E_{\kappa} = E_k +\omega_q~~;~~\vq = \vp - \vk$ and $\rho(p_0)$ is given by eqn. (\ref{rhosupfin}) and $\Delta = (\lambda/4\pi M)^2$. In terms of $s= (p_0-E_p)/E_p~~;~~R= M/E_p$ we find
\be \sum_{\kappa}|C_{\kappa}(\infty)|^2 =  \Delta\,R^2 \,\overline{Z}_{\phi}\,\int^{\infty}_{0} \Bigg[ \frac{2+s}{R^2+2s+s^2}\Bigg]\,  s^{2\Delta-1} \,  {ds}\,, \label{sumkapa} \ee notice that naively setting $\Delta=0$ in $s^{2\Delta-1}$ in the integrand in (\ref{sumkapa}) leads to an infrared divergence. It is precisely this anomalous dimension that renders the integral finite and $\propto 1/\Delta$ thereby cancelling the $\Delta$ in the prefactor. This is seen as follows: writing $\int^\infty_0 \cdots \, ds = \int^1_0 \cdots \, ds + \int^\infty_1 \cdots \, ds$ and in the first integral replace
\be \Bigg[ \frac{2+s}{R^2+2s+s^2}\Bigg] = \frac{2}{R^2} + \frac{s}{R^2} \Bigg[ \frac{R^2-4 -2s}{R^2+2s+s^2}\Bigg]\,, \label{repla}\ee  the first term on the right hand side of  (\ref{repla}) when input in the first integral ($\int^1_0 \cdots \, ds $) yields $1/(R^2\Delta)$, finally yielding
\be \sum_{\kappa}|C_{\kappa}(\infty)|^2 = \overline{Z}_\phi \,\Big[1 + \Delta\,     \mathcal{I}[R;\Delta] \Big] \,,\label{C2kapfina}\ee
where
\be \mathcal{I}[R;\Delta] =  \int^1_0 \Bigg[ \frac{R^2-4 -2s}{R^2+2s+s^2}\Bigg]\,s^{2\Delta}\,ds + R^2\, \int^{\infty}_1 \Bigg[ \frac{2+s}{R^2+2s+s^2}\Bigg]\,  s^{2\Delta-1}   {ds} \,. \label{Iints}\ee
With $\Gamma[1-2\Delta]\,\sin\big[\frac{\pi}{2}(1-2\Delta) \big] = 1+ 2\Delta\,\gamma_E + \mathcal{O}(\Delta^2)$ and from eqn. (\ref{zfi},\ref{Cfi2}) $ \mathcal{Z}_\phi=1- 2\Delta \gamma_E -\Delta \, \mathcal{I}[R;0] + \mathcal{O}(\Delta^2)$ it follows that $\overline{Z}_\phi = 1  -\Delta \,\mathcal{I}[R;0] + \mathcal{O}(\Delta^2)$. Since $\mathcal{I}[R;\Delta]$ is infrared finite, to lowest order we can replace   $\mathcal{I}[R;\Delta]\simeq \mathcal{I}[R;0]+\mathcal{O}(\Delta)$ inside the bracket in eqn. (\ref{C2kapfina}),  hence,  neglecting consistently terms of $\mathcal{O}(\Delta^2) \simeq H^4_I$ and higher,  we find
\be  \sum_{\kappa}|C_{\kappa}(\infty)|^2 = 1 + \mathcal{O}(\Delta^2\simeq H^4_I)\,. \label{fulfiuni}\ee therefore unitarity is fulfilled consistently up   order $\mathcal{O}(H^4_I)$ that we have considered. It now becomes clear that the non-perturbative dynamical renormalization group resummation yielding the anomalous dimension is precisely the mechanism by which unitarity is fulfilled.
The extension to the renormalizable case with scalar Yukawa coupling is  straightforward with a similar result up to   order $\mathcal{O}(H^4_I)$ that we considered.

\textbf{No infrared divergences:} In the cases where there are no infrared divergences, such as that of axion like particle couplings, the asymptotic long time dynamics follows from eqn. (\ref{ampslt}) with $\Delta =0$, namely
\be C_A(t) = \mathcal{Z}^{1/2}_A\, e^{-i\delta E_{\infty}\,t}~~;~~ \mathcal{Z}_A =  e^{- {z_A}}  \,,\label{ampslt2}\ee with (see eqns. (\ref{zA},\ref{wavefun1}))
\be  z_A = 2 \int^{\infty}_{-\infty}  \, \,\frac{\rho(p_0)}{(E_A-p_0)^2}\,dp_0 \,. \label{zaNIR} \ee Inserting these expressions into eqn. (\ref{modC2}),  absorbing the phase $\Delta E_{\infty}$ into a renormalization of the single particle frequencies and carrying out the time integrals,  we now find
\be |C_{\kappa}(t)|^2 = \mathcal{Z}_A\, |\bra{A}H_I(0)\ket{\kappa}|^2 \Bigg[\frac{1-\cos[(E_\kappa-E_A)t]}{(E_\kappa-E_A)^2} \Bigg]   \,,  \label{modC2NIR}\ee yielding
\be \sum_{\kappa}  |C_{\kappa}(t)|^2 = 2\,\mathcal{Z}_A\,\int  \rho(p_0) \,  \Bigg[\frac{1-\cos[(p_0-E_A)t]}{(p_0-E_A)^2} \Bigg] \, dp_0\,.  \label{NIRsum} \ee Since in the cases in which there are no infrared divergences, as $p_0 \rightarrow E_A$
\be \rho(p_0) \simeq (p_0-E_A)^n ~~;~~ n\geq 2 \,, \label{NIRrho}\ee the oscillatory contribution in (\ref{NIRsum}) averages out in the long time limit yielding the asymptotic behavior as $t\rightarrow \infty$
\be \sum_{\kappa}  |C_{\kappa}(\infty)|^2 = 2\,\mathcal{Z}_A\,\int    \,  \frac{\rho(p_0)}{(p_0-E_A)^2}   \, dp_0\,.  \label{NIRsumasy} \ee The unitarity relation (\ref{unitarity1}) at asymptotically long time  becomes,
\be \mathcal{Z}_A \Bigg[1+ 2\, \int\frac{\rho(p_0)}{(p_0-E_A)^2}   \, dp_0\, \Bigg] =1\,. \label{NIRunit}\ee  with
 \be \mathcal{Z}_A = e^{-z_A} \simeq 1-z_A +\cdots = 1- 2 \,\int \,    \frac{\rho(p_0)}{(p_0-E_A)^2}   \, dp_0\,+ \mathcal{O}(H^4_I) \,,\label{NIRunit2} \ee it is clear that the unitarity relation (\ref{NIRunit}) is fulfilled up to $\mathcal{O}(H^4_I)$ which is consistent with the order   that we have considered.

\subsection{The entangled dressing cloud:}\label{subsec:cloud}
Focusing on the super  renormalizable case, with an obvious extension to the renormalizable case, the states $\ket{A} = \ket{1^{\phi}_{\vp}}$ and $\ket{\kappa}= \ket{1^{\phi}_{\vk}\,;\,1^{\chi}_{\vq}}$ with $E_{\kappa} = E_k +q$. Denoting the coefficients $C_{A}(t) \equiv C_{p}(t) ~;~ C_{\kappa}(t) \equiv C_{\vk,\vq}(t)$, the time evolved state (in the interaction picture) is
\be \ket{\Psi(t)} =  C_{p}(t) \,\ket{1^{\phi}_{\vp}} + \sum_{\vq,\vk} C_{\vk,\vq}(t)\, \ket{1^{\phi}_{\vk}\,;\,1^{\chi}_{\vq}}\,,  \label{psioft} \ee
where
\be C_{p}(t) = e^{-i\delta E_\infty\, t} \, \Big[E_p\,t \Big]^{-\Delta}\, {\mathcal{Z}^{1/2}_\phi} \,, \label{cpofti}\ee and the coefficients $C_{\vk,\vq}(t)$ obtained from eqn.(\ref{Ckapasol}).
The asymptotic state after the probability to remain in the initial state has vanished is given by (in the interaction picture)
\be \ket{\Psi(\infty)}= \sum_{\vk,\vq}C_{\vk;\vq}(\infty) \ket{1^{\phi}_{\vk}\,;\,1^{\chi}_{\vp-\vk}}\,,\label{cloud} \ee
with (see eqn. (\ref{Ckapinfty}))
\be |C_{\vk;\vq}(\infty)|^2 = 2 \, \mathcal{Z}_{\phi}\, \Bigg[ \frac{|\bra{1^{\phi}_{\vp}}H_I(0)\ket{1^{\phi}_{\vk}\,;\,1^{\chi}_{\vq}}|^2}{(E_k +q-E_p)^2}\Bigg]\, \Big[\frac{E_k +q-E_p}{E_p} \Big]^{2\Delta}\, \Gamma[1-2\Delta]\,\sin\big[\frac{\pi}{2}(1-2\Delta) \big]~~;~~\vq = \vp-\vk\,, \label{Cfiasy} \ee where the corresponding matrix elements are given by eqn. (\ref{mtxel1}), and from eqn. (\ref{fulfiuni})
\be \sum_{\vk,\vq} |C_{\vk;\vq}(\infty)|^2 =1+ \mathcal{O}(H^4_I)\,. \label{unita2} \ee

It is important to compare the time evolved state (\ref{psioft}) with previous studies. In ref.\cite{finites} the state dressed by soft massless quanta was obtained up to first order in time-ordered perturbation theory (see eqn. (5) in ref.\cite{finites}), whereas the state $\ket{\Psi(t)}$ (\ref{psioft}) describes a non-perturbative resummation of the perturbative series, as is evident in the anomalous dimension.

Furthermore, the dressed states considered in references\cite{chung,kibble,kulish,zell,tomaras} are built from a coherent state of photons, which are very different from the state $\ket{\Psi(t)}$ which in the long time limit  is a superposition of   single charged and a single massless particle states, and the probabilities include wave function renormalization constants.

This is an entangled state of   the charged particle $\phi$   and the cloud of massless quanta $\chi$. At asymptotically long time, the probability of finding a   $\phi_{\vk},\chi_{\vq}$ pair is
$|C_{\vk,\vq}(\infty)|^2$ and by unitarity $\sum_{\vq,\vk} |C_{\vk,\vq}(\infty)|^2 =1$ as explicitly shown in eqn. (\ref{fulfiuni}). The density matrix
\be \wp =  \ket{\Psi(\infty)}\bra{\Psi(\infty)} \label{densmtx}\ee describes a pure state with $\mathrm{Tr} \wp =1$.

However, consider that this asymptotic state is measured by a detector that is only sensitive to the charge of the $\phi$ field, but insensitive to the charge neutral massless quanta of the $\chi$ field. Such measurement amounts to tracing the density matrix over the unobserved $\chi$ field  yielding the mixed state described by the reduced density matrix
\be \mathrm{Tr}_{\chi} \wp = \sum_{\vk}|C_{\vp;\vk}(\infty)|^2 \, \ket{1^{\phi}_{\vk}}\bra{1^{\phi}_{\vk}} \,, \label{mixstat}\ee from which we interpret $|C_{\vp,\vk}(\infty)|^2$ as the distribution function of the charged fields in the asymptotic state.

Although this discussion has focused on the asymptotic state, it can be extended to include the full time evolution of the state $\ket{\Psi(t)}$. Tracing over the unobserved $\chi$ states the reduced density matrix at any given time is
\be \mathrm{Tr}_{\chi} \wp (t)= |C_p(t)|^2 \,  \ket{1^{\phi}_{\vp}} \bra{1^{\phi}_{\vp}}+\sum_{\vk}|C_{\vp;\vk}(t)|^2 \, \ket{1^{\phi}_{\vk}}\bra{1^{\phi}_{\vk}}\,.
\label{rhoredoft} \ee
This mixed state features a von Neumann entropy
\be S_{\phi}(t) = -|C_p(t)|^2 \,\ln\big(|C_p(t)|^2  \big)-\sum_{\vk} |C_{\vp;\vk}(t)|^2\,\ln\Big( |C_{\vp;\vk}(t)|^2\Big)\,, \label{vNent}\ee since $C_{p}(0)=1~;~C_{\vp;\vk}(0)=0$,
it follows that $S_{\phi}(0)=0$. The time evolution of the entropy is completely determined by the (DRM) equations (\ref{Ckapasol},\ref{intdiff})  and in the cases with infrared divergences $C_{p}(\infty)=0~;~C_{\vp;\vk}(\infty)\neq0$, asymptotically at long time
 \be S_{\phi}(\infty) = -\sum_{\vk} |C_{\vp;\vk}(\infty)|^2\,\ln\Big( |C_{\vp;\vk}(\infty)|^2\Big) > 0 \,, \label{vNentasy}\ee with the probabilities $|C_{\vp;\vk}(\infty)|^2$ obeying the unitarity condition (\ref{unita2}). Unitary time evolution entails a flow of information from the initial single particle state to the asymptotic entangled two particle state with a concomitant growth of the entanglement entropy whose  time evolution   is completely determined by the (DRM) equations (\ref{Ckapasol},\ref{intdiff}).

 The entanglement entropy resulting from the correlations between hard charged particles and soft photons in QED was studied in ref.\cite{tomaras} within the context of the coherent dressed states proposed in ref.\cite{kulish}. As mentioned above these states are very different from the dressed state obtained from the unitary time evolution of the initial single particle state by the (DRM), thus preventing a meaningful comparison.

The entanglement entropy (\ref{vNentasy}) is infrared finite, although the coefficients $|C_{\vp;\vk}(\infty)|^2$ feature the infrared enhancement near threshold $E_{k}+q \rightarrow E_p$ exhibited by their denominators in   (\ref{Cfiasy}),  it  is compensated by the power law with anomalous dimension in the numerator. An integral within a small region in which the denominator vanishes is rendered finite by the anomalous dimension. Indeed, as discussed in the previous section, it is the numerator with the anomalous dimension in the coefficients (\ref{Ckapinfty}) that ultimately leads to an infrared finite integral of $|C_{\vp;\vk}(\infty)|^2$ and the fulfillment of unitarity, eqn. (\ref{fulfiuni}).

The calculation of the  entanglement entropy either (\ref{vNent}) or its asymptotic form (\ref{vNentasy}) is complicated by the logarithms and does not seem  a priori to yield a  useful quantity since it   depends non only on the anomalous dimension but also on the couplings, the volume\footnote{A similar volume dependence has been discussed in ref.\cite{tomaras}.} and  the ultraviolet aspects of the theory through the wave function renormalization constant $\mathcal{Z}$.    While the growth of entropy and information flow from the initial state to the asymptotic multiparticle state as a consequence of unitary time evolution and its dependence on the anomalous dimension are interesting conceptually, it remains to be understood if it provides any observational consequence.

\section{Discussion:}

\vspace{1mm}

\textbf{Scaling behavior and renormalization group invariance:} In the  cases in which there is an infrared divergence at threshold the survival probability  at long time is given by
\be |C_A(t)|^2 = \mathcal{Z}_A \, \Big[E_A t\Big]^{-2\Delta}\,. \ee This scaling behavior  can be written in a manifestly renormalization group invariant form as
\be |C_A(t)|^2 = \mathcal{Z}_A[\mu]  \, \big[\mu t\big]^{-2\Delta}~;~~ \mathcal{Z}_A[\mu]= \mathcal{Z}_A  \Big[\frac{E_A}{\mu}\Big]^{-2\Delta} \,,\label{probrginv}\ee so that $|C_A(t)|$ is independent of the renormalization scale $\mu$, namely
\be \frac{\partial |C_A(t)|^2}{\partial \ln[\mu]} = 0 \Rightarrow \frac{\partial \ln \mathcal{Z}_A[\mu]}{\partial \ln[\mu]} = 2\Delta ~~;~~ \mathcal{Z}_A[E_A]=\mathcal{Z}_A \,. \label{rginva}\ee

The renormalization group invariance of the above result can also be made explicit by noting  that we can also write the integral in the first term in $I_1(T)$ (\ref{I1sepa}) as $\int^{s_0}_0 + \int^{1}_{s_0} $ with $s_0 = \mu/E_p$, and $\mu$ an arbitrary renormalization scale. Obviously the result is independent of the scale $\mu$ and the long time limit (\ref{gamafilt}) would become
\be   \gamma(t)_{~~ \overrightarrow{t\longrightarrow \infty}~~}  2 \,\Delta\, \ln\big[\mu t \big]+ z_{\phi}[\mu]\,. \label{gamafiltmu} \ee Since $\gamma(t)$ is independent of the arbitrary scale $\mu$ it obeys the renormalization group equation
\be \mu \frac{d\,\gamma(t)}{d\mu} = 0 \,, \label{rgeqngama}\ee  the solution (\ref{Cfi2}) corresponds to   $\mu=E_p$.

\vspace{1mm}

\textbf{Exponential vs. power law decay:} Instead of  the model described by the Lagrangian density (\ref{lagsuper}) with two fields, let us consider, for example the case of three fields $\phi_1,\phi_2,\chi$ with $\chi$ a massless field and both $\phi_1,\phi_2$ are charged and massive with $M_1 > M_2$ and a cubic interaction among all fields, $\mathcal{L}= \lambda \, \phi^\dagger_1 \, \phi_2 \,\chi +\mathrm{h.c.}$. In this case the heavier field $\phi_1$ can decay into the $\phi_2,\chi$, hence the $\phi_1$ single particle pole is now embedded in the two-particle continuum with threshold at $M_2< M_1$. The survival probability for a single $\phi_1$  particle state of momentum $\vp$  decays in time in the long time limit as $e^{-\Gamma_p \,t}$ with $\Gamma_p = 2\pi\,  \rho(p_0 = E^{(1)}_{p})$ . As $M_2 \rightarrow M_1$ from below, the decay rate $\Gamma_p \rightarrow 0$ as now the threshold coincides with the position of the mass shell of $\phi_1$ and the spectral density vanishes at threshold by kinematics. This is the case in which infrared singularities emerge when the spectral density vanishes linearly at threshold. In this case the asymptotic long time limit is determined by the subleading secular terms that do not grow linearly in time, but as described above, only logarithmically, and as $M_2$ becomes larger than $M_1$ the single particle $\phi_1$ pole moves below the multiparticle threshold, it is now an isolated pole below the continuum describing a stable particle.

\vspace{1mm}

\textbf{Infrared dressing and the S-matrix:} In this article we focused on studying the dynamics of dressing by soft quanta directly in real time in model quantum field theories that feature infrared divergences akin to those in gauge theories. This is undoubtedly only a first step,  and of much more limited scope than addressing infrared aspects in S-matrix elements between asymptotic states in gauge theories. While a direct extrapolation of our results to the understanding or resolution of these divergences in S-matrix elements   must await a deeper study, we can comment on some \emph{possible} implications. To begin with, the S-matrix considers the time evolution of states prepared in the infinite past towards the infinite future, hence it is an infinite time limit of the  finite time analysis presented here. As we have shown, the amplitude of the single particle state vanishes as a power law with anomalous dimension in this limit, this is in agreement with the  vanishing of the on-shell wave function renormalization as a consequence of infrared divergences, namely the vanishing of the amplitude of the single particle ``pole''. Therefore, even when an initial single particle state is ``prepared'' in the infinite past, it dresses itself with soft quanta becoming the asymptotic entangled state given for example by (\ref{cloud}) with the coefficients obeying the unitarity condition (\ref{unita2}) a result of unitary time evolution as expressed by the sum rule (\ref{unistat}). It is then this entangled, multiparticle state that should be considered as the ``in'' state   and also describes the asymptotic ``out'' states in the S-matrix calculation of  a cross section or transition rate. Therefore, an assessment of the infrared finiteness of S-matrix elements between asymptotic states entails a calculation of the scattering processes not in terms of single particles, but in terms of the entangled multiparticle states of the form (\ref{cloud}). An analysis along these lines was originally presented in refs.\cite{chung,kibble} but with dressed states as coherent states, which are very different from the states (\ref{cloud}) as discussed above. Scattering of ``Kulish-Faddeev''\cite{kulish}  dressed states has been considered in ref.\cite{strominger1}, again, such states are strikingly different from the multiparticle state (\ref{cloud}) which has been obtained directly from the real time evolution and whose amplitude satisfies the sum rule (\ref{unita2}), a direct consequence of unitary time evolution. A challenging open question is how to incorporate the non-perturbative resummation that evolves the single particle state into the dressed entangled state (\ref{cloud}) consistently with Feynman calculus ubiquitous in S-matrix calculations.

Therefore,   the infrared finiteness of the S-matrix based on the dressed states (\ref{cloud}) whose spectral properties feature the anomalous dimensions associated with the non-perturbative resummation of infrared emission and absorption, remains an open question which undoubtedly  merits further and deeper study  well beyond the limited scope of this article.

\vspace{1mm}

\textbf{Loop corrections to the mass of the light field:} In this article we focused on the infrared aspects associated with the emission and absorption of a massless scalar particle which are akin to those in gauge theories. One of our motivations is to learn how to describe these processes in real time with a view towards a cosmological setting as a potential mechanism of production of ultra light dark matter. In a non-gauge theory the masslessness of the scalar field must be protected by some symmetry, for example the scalar field in our examples could be a Goldstone boson associated with a spontaneously broken symmetry beyond the standard model or an axion-like \emph{scalar} pseudo-Goldstone boson. In absence of protecting symmetries radiative corrections may induce a non-vanishing (and perhaps large) mass and such symmetry should be responsible for the near masslessness of such an ultra light dark matter candidate. While our results describe the dynamics of dressing and is of fundamental character, their applicability must be carefully considered in particular cases by assessing whether radiative corrections (higher order loop corrections) induce a large mass invalidating the results based on the masslessness of the scalar particle.

\section{Conclusions and further questions:}

In this article we studied the infrared aspects of the  dressing dynamics of charged states by the emission and absorption of massless neutral quanta directly in time, specifically in non-gauge theories. While motivated by possible cosmological implications for production of ultra-light axion-like particles, and focused on the time evolution of initial single particle states, our study provides a complementary exploration of infrared phenomena ubiquitous in the S-matrix formulation of gauge theories, and possibly of infrared phenomena in gravity. We have considered super renormalizable and renormalizable theories that while featuring very different ultraviolet behavior, nonetheless share similar infrared behavior. Infrared singularities in these theories arise as a consequence of the charged particle mass shell merging with  the beginning of a multiparticle branch cut in the charged particle self-energy and are, therefore, akin to infrared divergences in gauge theories. We map this infrared divergence into similar divergences of an Euclidean critical theory and implement a renormalization group resummation of the propagator yielding scaling behavior near threshold. This  translates into a survival probability of the charged single particle state that decays as a power law in time with an anomalous dimension, namely $t^{-\Delta}$. We introduced a dynamical resummation method that extends the dynamical renormalization group and obtain the time dependent amplitudes of the single charged particle state, and the excited multiparticle states. This method is manifestly unitary and yields the survival probability directly in time. It clearly reveals that infrared dynamics arises when the spectral density vanishes linearly at threshold and yields the power law decay of the survival probability $t^{-\Delta}$ explicitly relating the anomalous dimension $\Delta$ to the slope of the spectral density at threshold. This behavior points to a certain universality in the sense that theories with very different ultraviolet behavior but with similar behavior of the spectral density near threshold feature similar power law decay   with anomalous dimensions.

The dynamical resummation method yields the unitary time evolution of the single charged particle state and explicitly shows that the dressed state is an  entangled state  of the charged field and massless quanta.

Tracing over the massless neutral quanta yields a reduced density matrix from which we extract the entanglement entropy at all time. Unitary time evolution entails an information flow from the initial single particle state with vanishing entropy to the asymptotic dressed state with an infrared finite  entropy   as a consequence of the anomalous dimension.

We find that effective field theories of massless axion-like particles coupled to charged fermions do not feature infrared divergences and provide a criterion generally valid for non-gauge theories up to one loop to determine if and when infrared divergences arise.

These results   lead to several questions that merit further study: \textbf{i)} how to extend the (DRM) to gauge theories consistently with gauge invariance, \textbf{ii)} how to combine the (DRM)  that describes the time evolution of initial states  with the S-matrix, which describes transition amplitudes from in-states prepared in the infinite past, to out states in the infinite future.  \textbf{iii)} aspects of coherence of dressed states have been addressed recently in refs.\cite{carney,zell,tomaras}, the (DRM) yields the dressed state as a function of time, it is very different from that in these references. While   the entanglement entropy is infrared finite, it depends not only on the anomalous dimension $\Delta$ but also on the ultraviolet behavior of the theory with no a priori direct relationship to observables.

Perhaps the results of our study could lead to further understanding of infrared effects in gauge theories and gravity and may provide a useful framework to study similar phenomena directly in time in cosmology.

\acknowledgements
L. C.   thanks Pok Man Tam and Hongbo Cai for useful discussions and acknowledges support through NSF grant PHY-1820760.  DB gratefully acknowledges support from the U.S. National Science Foundation through grant award NSF 2111743.

 \appendix

 \section{Renormalization group improved propagator.}\label{app:RG}

  In this appendix we argue that the infrared divergence and the emergence
of anomalous dimensions can be understood by establishing a parallel with
{\em static} critical phenomena  in
Euclidean space-time for $P^2 \simeq M^2_p$. We will focus the discussion on the super renormalizable example of section (\ref{subsec:super}) but it will be clear that the same analysis holds for the renormalizable case (\ref{nearpole}), and indeed for any other case in which the infrared divergence is manifest as a logarithmic multiparticle branch cut that originates at the position of the mass shell.  First let us analytically continue to Euclidean momenta
\be P^2 \rightarrow -P^2_E \,,\label{Peuc}\ee and introduce the dimensionless variable
\be \overline{P}^2 = \frac{P^2_E}{M^2_p}+1\,, \label{dimlessPE} \ee which maps the threshold region $P^2 \simeq M^2_p$ into the region $\overline{P}^2\simeq 0$. Near $P^2 \simeq M^2_p$ the Euclidean irreducible two-point vertex function in the super renormalizable theory given by (\ref{Gnearpole})
becomes
\be \Gamma_E(\overline{P}^2) = -G^{-1}(P^2_E) = M^2_p \overline{P}^2 \Big[1 - g^2\,\ln\big[ \overline{P}^2 \big] \Big] ~~;~~ g  = \frac{\lambda}{4\pi M_R}\,,    \label{vereuc}\ee the threshold infrared divergence is now cast as an infrared divergence in a critical (massless) theory\cite{amit} as $\overline{P}^2 \rightarrow 0$. Let us introduce  a wavefunction renormalization constant
\be Z(\overline{\mu}) = 1 + g^2 \ln(\overline{\mu}^2) +\cdots \label{Zeuc}\ee where $\overline{\mu}$ is a (dimensionless) renormalization scale, and the renormalized vertex function
\be \Gamma_R(\overline{P}^2;\overline{\mu}^2) = Z(\overline{\mu})\,\Gamma_E(\overline{P}^2)\,,  \label{renGam}\ee with the condition
\be \Gamma_R(\overline{P}^2;\overline{\mu}^2)\big|_{\overline{P}^2=\overline{\mu}^2}= M^2_P\,\overline{\mu}^2 \,. \label{renpt}\ee
The \emph{bare} vertex function (\ref{vereuc}) is independent of the renormalization scale $\overline{\mu}$ which leads to the renormalization group equation
\be \Big[\overline{\mu} \frac{\partial}{\partial \overline{\mu}} - \eta\Big]\Gamma_R(\overline{P}^2;\overline{\mu}^2) = 0 ~~;~~ \eta = \frac{\partial \ln[Z](\overline{\mu})}{\partial \ln(\overline{\mu})} = 2\,g^2 \,. \label{REeqn}\ee By dimensional analysis, the renormalized vertex can now be written as
\be \Gamma_R(\overline{P}^2;\overline{\mu}^2) = M^2_P\,\overline{P}^2\,
\Phi\Big[\frac{\overline{P}}{\overline{\mu}} \Big] \,.\label{Phifu}\ee Using the renormalization group equation (\ref{REeqn}) and the boundary condition (\ref{renpt}),  we find that the dimensionless \emph{scaling} function $\Phi$ obeys
\be \Bigg[ \overline{P}\frac{\partial}{\partial \overline{P}} + \eta\Bigg] \Phi\Big[\frac{\overline{P}}{\overline{\mu}} \Big] = 0 ~~;~~ \Phi(1)=1 \,, \label{RGPhi} \ee with the solution
\be \Phi\Big[\frac{\overline{P}}{\overline{\mu}} \Big]  = \Big[\frac{\overline{P}}{\overline{\mu}} \Big]^{-\eta}\,,  \label{soluPhi}\ee yielding the renormalization group improved Euclidean propagator
\be G_R(\overline{P}^2) = -\frac{1}{M^2_P\,\overline{P}^2 \,\Big[\frac{\overline{P}^2}{\overline{\mu}^2} \Big]^{-g^2} }\,, \ee which upon analytic continuation back to Minkowski space-time yields (\ref{RGprop}) when taking the arbitrary (dimensionless) renormalization scale $\overline{\mu}=1$.

In this analysis the coupling has been considered as constant, namely a ``fixed point'' (not running with the renormalization group scale), because the infrared limit $\overline{P}_E \rightarrow 0$ actually corresponds to the momenta approaching the threshold value $P^2 \rightarrow M^2_p$ not an asymptotically large or small value as is envisaged in the usual running of the coupling under the usual renormalization group.

\end{document}